\documentclass[prl,twocolumn]{revtex4}

\usepackage{graphicx}
\usepackage{dcolumn}
\usepackage{bm}
\newcommand{\YSO}{Y$_2$SiO$_5$}
\newcommand{\del}{\partial}

\begin{document}

\title{Photon Echoes Produced by Switching Electric Fields}

\author{A. L. Alexander}
\author{J. J. Longdell}
\author{M. J. Sellars}
\author{N. B. Manson}
 \affiliation{Laser Physics Centre, Research School of Physical
    Sciences and Engineering, Australian National University,
    Canberra, ACT 0200, Australia.}
\begin{abstract}
  We demonstrate photon echoes in Eu$^{3+}$:Y$_{2}$SiO$_{5}$ by
  controlling the inhomogeneous broadening of the Eu$^{3+}$
  $^{7}$F$_{0}\leftrightarrow^{5}$D$_{0}$ optical transition.  This
  transition has a linear Stark shift and we induce inhomogeneous
  broadening by applying an external electric field gradient.  After
  optical excitation, reversing the polarity of the field rephases the
  ensemble, resulting in a photon echo.  This is the first
  demonstration of such a photon echo and its application as a quantum
  memory is discussed.
\end{abstract}


\maketitle

In the emerging area of quantum information science the ability to
store and recall quantum states of light it is highly desirable.  To
date all proposals to achieve this rely on the mapping of light states
onto the states of atom-like systems. In spite of the significant
effort that has been directed towards this goal, high fidelity reversible
mapping of the state of a light field onto atom-like systems has not
yet been demonstrated.

The use of atomic ensembles for quantum memory applications is
attractive because it allows strong coupling without the need for the
high finesse cavities of earlier proposals \cite{cira97}.  Much
attention has be been given to using the phenomena of
electromagnetically induced transparency (EIT) to build quantum
memories using atomic ensembles \cite{flei00,flei02}.  EIT has led to
some dramatic experimental results including ultra-slow group
velocities and light storage in trapped atomic systems
\cite{hau99,liu01}, atomic vapors \cite{kash99,budk99,phil01} and the
solid state \cite{turu02,long05}. While EIT can lead to very slow
group velocities it has proven difficult to achieve delays with large
time-bandwidth products and without high losses \cite{mats05}. Low
time-bandwidth products for delays make high fidelity light storage
using EIT difficult. This is because the time-bandwidth product is a
measure of how-many distinct pulses fit inside the delay medium and
for high fidelity storage the entire pulse must remain inside the
medium as the group velocity is slowly reduced to zero.  
In this work we explore an alternative ensemble based quantum memory
on photon echoes.


Coherent manipulation and storage of classical light states using
photon echo techniques dates back to the 1980's \cite{moss82}.  In
contrast to EIT based techniques where the required slow group
velocities are directly linked to narrow bandwidths, recent photon
echo based experiments have demonstrated the ability to store
thousands of pulses \cite{lin95} and do signal processing at gigahertz
bandwidths \cite{merk04}. However there is no way currently known to
use a standard photon echo, where the rephasing of the atomic
coherence is achieved with an optical $\pi$-pulse, in a quantum
memory.

In 2001 Moiseev and Kr\"oll \cite{mois01} published a proposal for a
quantum memory based on modified photon echos. Instead of rephasing
the atomic coherence with intense optical pulses, it used reversible
inhomogeneous broadening. Here we report the first demonstration of
photon echoes produced by reversible inhomogeneous broadening. In this
case this is achieved using optical centers in a solid which have a
linear stark shift and macroscopic electric field gradients.

A quantum memory based on controlled inhomogeneous broadening can be
understood as follows: consider a coherent light pulse entering a
medium of two-level atoms. The Hamiltonian for the system is of the
form:
\begin{equation}
    H = H_{\textrm{field}} + \sum_n \omega_n
  \sigma_n^z + d(\sigma_{n}^+E(z_n,t)+\sigma_{n}^-E^{\dagger}(z_n,t))
\end{equation}
Here the $\sigma$ are Pauli operators and $d$ the transition dipole
moment. We assume that the area of the incoming pulse is small
and that each individual atom is never driven far from its ground
state, enabling  the approximation $\sigma_z=-1/2$. Working
in this small-pulse regime linearizes the
equations of motion allowing simple analytic expressions to be
derived. It is also the regime of interest for a
quantum memory. 

The treatment here is semi-classical but because
the equations of motion are linear the results should carry straight
over to a fully quantum analysis. Treating the atoms as a continuous
field $\sigma_n^-(t)\rightarrow \varphi(\delta,z,t)\,dz\,d\delta$ one
obtains the following equations of motion:
\begin{eqnarray}
  \label{eq:maxbloch}
  (-\frac{\del^2}{\del z^2} + \frac{1}{c^2}\frac{\del^2}{\del
    t^2})E(z,t) = \eta\int d\delta \, g(\delta,z)\, \frac{\del^2}{ \del
    t^2}\varphi(\delta,z,t)\\
\frac{\del}{\del t} \varphi(\delta,z,t) = -i (\omega_0+\delta) \varphi(\delta,z,t) +d\ E(z,t)
\end{eqnarray}
Here $g(\delta,z)$ describes the atom density as a function of
detuning and position.  Changing
variables by setting 
\begin{eqnarray}
  E(z,t) &=& E_f(z,t)\exp(i(kz-\omega_0 t))\\
\varphi(\delta,z,t)& =& \alpha(\delta,z,t)\exp(i(kz-\omega_0 t))
\end{eqnarray}
 and
assuming that the amplitudes $E_f$ and $\alpha$ are slowly varying functions of
both $z$ and $t$. We obtain
\begin{eqnarray}
  2ik\left(\del_z - \frac{1}{c}\del_t\right)E_f &=& \eta\omega_0^2 \int 
d\delta\, g(\delta,z) \alpha(\delta,z,t)\nonumber\\
\label{eq:forward}
\del_t \alpha(\delta,z,t)& = &-i\delta \alpha(\delta,z,t) + E_f(z,t)
\end{eqnarray}
If we narrow our focus briefly to the situation where we have a
spatially homogeneous system with large inhomogeneous broadening
($g(\delta,z)=1$) these have the analytical
solution \cite{cris70}
\begin{eqnarray}
  \label{eq:solns}
  E_f(z,t) z&=& e^{-\eta z} E_f(0,t-z/c)\\
\alpha(\delta,z,t) &=& e^{(i\delta/c - \eta)z}\int_{-\infty}^t d\tau E_f(0,\tau)e^{i\delta \tau}
\end{eqnarray}
As it propagates, the optical pulse exponentially decays as the atoms
absorb the energy. Because of the inhomogeneous broadening the
macroscopic coherence of the atomic ensemble also decays. However the
process should not be seen as dissipative, each individual atom has
not lost any coherence and the absorption process is reversible as
described below. For the experiments we performed the the detuning is
a function of position and the decay of the field is not exponential.
The different spectral components of the input field are
absorbed at different positions. However as long light pulse is
could be totally absorbed the argument for how the
memory works is unchanged.

Defining  backwards propagating atomic ($\beta$) and optical ($E_b$) fields
 via
\begin{eqnarray}
  E(z,t) &=& -E_b(z,t)\exp(i(-kz-\omega_0 t)) \\
\varphi(\delta,z,t)& =& \beta(\delta,z,t)\exp(i(-kz-\omega_0 t))
\end{eqnarray}
one arrives at the equations of motion
\begin{eqnarray}
  2ik\left(\del_z +\frac{1}{c}\del_t\right)E_b(z,t) = \eta\omega_0^2 \int 
d \delta\, g(\delta,z)\beta(\delta,z,t)\nonumber\\
\label{eq:backward}
\del_t \beta(\delta,z,t) = -i\delta \beta(\delta,z,t) - E_b(z,t)
\end{eqnarray}
Comparing Eqns.~(\ref{eq:backward}) with the time reversed versions of
(\ref{eq:forward}) it can be seen that the two coincide if the sign of
$\delta$ is reversed.  Thus all that is required to make the pulse
come out again as a time reversed copy of itself is to flip the
detunings of the atoms and at the same time to apply an phase matching
operation such that the value of $\beta$ after the operation is equal
to the value that $\alpha$ was before the operation.
As $\beta(\delta,z,t) = \alpha(\delta,z,t) \exp(i2kz)$ this phase
matching operation is a position dependent phase shift for the atomic
states. This can be achieved by driving from the excited state to an
auxiliary ground state with a $\pi$ pulse and driving back up again
with another $\pi$ pulse such that the wavevector difference is $2k$.
These two $\pi$ pulses can be separated in time and between them the
coherence is stored in the hyperfine transitions.  Coherence times of
many seconds have been demonstrated for hyperfine transitions
\cite{frav05} in rare earth ion doped systems.

In \cite{mois01} it was proposed that the reversible inhomogeneous
broadening be achieved using Doppler broadening in an atomic gas.
Since then using impurity ions to achieve this controlled inhomogeneous
broadening has been proposed \cite{mois03} and independent of this work
the use of electric field gradients has also been proposed  \cite{nils05}.
A theoretical treatment of the general case has also appeared \cite{krau05}.

In order for a quantum memory based on controlled inhomogeneous
broadening to be practical it is necessary for the induced broadening
to be larger than the unbroadened linewidth of the transition. A further
requirement is that  the field
polarity  be switched in a time short compared to the inverse of this
linewidth.  Here we show that these conditions can be achieved in
rare earth ion doped systems by successfully demonstrating a {\it
  Stark echo}.



\begin{figure}[!ht]
  \includegraphics[height=4cm]{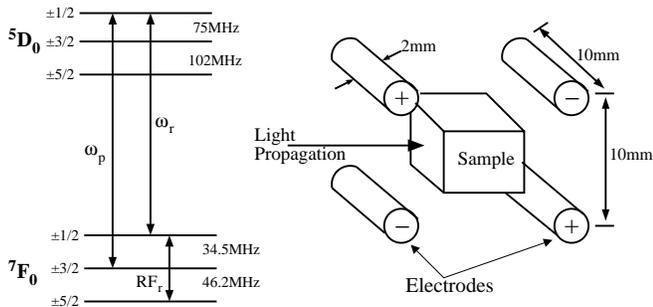}
  \caption{\label{electrodes}Energy level diagram of Eu$^{3+}$:\YSO\ 
and the arrangement of the electrodes around the sample. The
    experiment was carried out on the transition labelled $\omega_{p}$.  
$\omega_{r}$ and $RF_{r}$ were used to optically pump the
    desired ions into the $\pm$3/2 hyperfine state.}
\end{figure}

The optical transition used in this experiment was the $^{7}$F$_{0}
\rightarrow$ $^{5}$D$_{0}$ in $^{151}$Eu, at 579.879~nm in 0.1 at\%
Eu$^{3+}$:Y$_{2}$SiO$_{5}$. FIG.~\ref{electrodes} shows the hyperfine
structure of the two electronic singlet states.  The transition was
excited with linearly polarized light propagating along the C$_2$ axis
of the crystal, with the polarization chosen to maximize the
absorption.  The length of the crystal in the direction of propagation
was 4~mm. The crystal was cooled to below 4~K in a liquid helium bath
cryostat. A quadrupole electric field was applied to the sample using
four 10~mm long, 2~mm diameter rods in a quadrupolar arrangement as
shown in FIG.~\ref{electrodes}. Two  amplifiers with 1~MHz bandwidth
supplied the voltage across the electrodes.  These amplifiers had two
opposite polarity outputs and voltage rails of $\pm$35~V. This
configuration provided an electric field that varied linearly across
the sample in the direction of light propagation with a maximum field
gradient of approximately 300~Vcm$^{-2}$.

The optical setup was essentially the same as in previous work
\cite{pryd00,tomog,long04}. A highly stabilized dye laser was
used with an established stability of better than 200~Hz over
timescales of 0.2~s. The light incident on the sample was gated
with two acousto-optic modulators (AOMs) in series. These allowed
pulses with an arbitrary amplitude and phase envelope to be
applied to the sample. A Mach-Zehnder interferometer arrangement
with the AOMs and sample in one arm was employed to enable
heterodyne detection of the coherent emission from the sample. The
overall frequency shift introduced by the AOMs was 51~MHz.  The
intensity of the beat signal was detected with a photo-diode.

The linear Stark shift for the $^{7}$F$_{0} \rightarrow$ $^{5}$D$_{0}$
transition in Eu$^{3+}$:\YSO\ has not been reported but from a study
in YAlO$_{3}$ it is expected to be of the order of 35~kHzVcm$^{-1}$
\cite{meix92}. With the current experimental setup the anticipated
Stark-induced spectral broadening was therefore 2~MHz.  Although this
broadening is large compared to the 122~Hz homogeneous linewidth of
the optical transition it is significantly smaller than the
inhomogeneous linewidth of our sample which was 3~GHz.  To create an
optical feature which was narrow compared to the induced broadening,
the same optical pumping procedure as used in \cite{tomog,long04} was
employed. This consisted of burning a relatively wide ($\approx$3~MHz)
spectral hole in the absorption line by scanning the laser frequency.
A narrow anti-hole was placed in the middle of this region by applying
RF excitation at 80.7~MHz as well as light at a different frequency.
This frequency was given by the combination of ground and excited
state hyperfine splittings as shown in FIG.~\ref{electrodes}.  The
spectral width of the anti-hole was then reduced by optically pumping,
out of resonance, ions more than 12.5~kHz from the center frequency of
the anti-hole.  The peak absorption of the feature was approximately 
40\%.

\begin{figure}[!ht]
\includegraphics[width=0.35\textwidth]{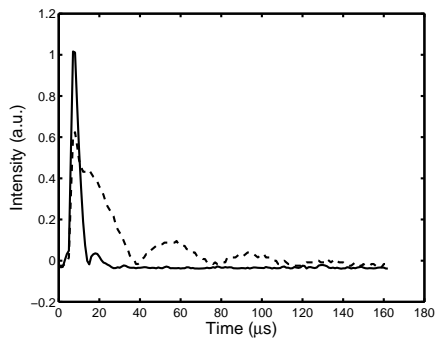}
\caption{\label{FID}Broadening of a narrow spectral feature with the 
electric field gradient.  The dotted line shows the FID from a 
prepared 25~kHz wide feature.  The solid line shows how this 
feature is broadened (and hence its FID is shortened) by the 
application of the field gradient.  The solid trace was taken with 
$\pm$4.5~V on the electrodes.}
\end{figure}

Figure~\ref{FID} (dotted) shows the free induction decay (FID) resulting from
excitation of the spectral feature created as described
above, with a 3~$\mu$s long optical pulse at $\omega_p$. The 
coherent emission was found to be consistent with that
from a 25~kHz wide feature with a top hat profile.

In order to determine the degree of induced broadening, the FID measurement was
repeated with an electric field gradient applied after the
creation of the spectral feature. By measuring the length of FIDs as the
voltage was increased, we estimate the rate of induced broadening to 
be 42~kHzV$^{-1}$.

To observe an echo from a feature broadened by applying 25~V to the 
electrodes, we excited the feature using a 1~$\mu$s optical pulse.  
The polarity of the field was reversed after a time $\tau$, and 
after a further delay of $\tau$ the echo was observed.

\begin{figure}[!ht]
\includegraphics[width=0.33\textwidth]{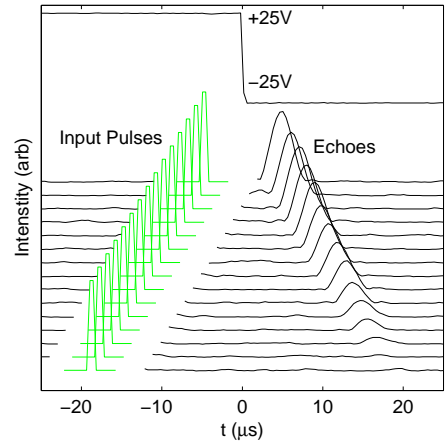}
\caption{\label{Stark_echo}Stark echoes produced by switching the
  polarity of the electric field.  The top trace is the voltage on the
  electrodes and the lower traces show the echoes.  The signal from
  the transmitted input pulses saturated the detector an so are only
  shown schematically. The input pulses were 1~$\mu$s square shapped
  pulses and a ratio of 60~dB between the input pulse and the largest
  echo was measured using a neutral density filter after the sample.
  The output pulses are broader in time than in input pulses because
  of the finite bandwidth of the memory.
  The voltages on the electrodes were $\pm$25~V.}
\end{figure}

\begin{figure}[!ht]
\includegraphics[width=0.35\textwidth]{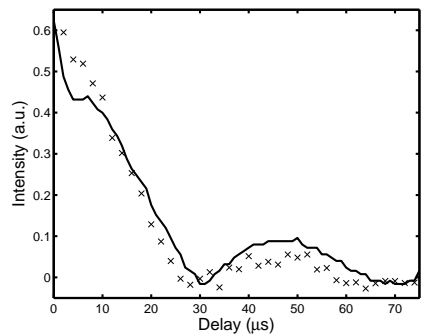}
\caption{\label{coherence}The solid line represents the FID of a
  3~$\mu$s pulse absorbed over a 25~kHz wide spectral feature.  The
  crosses show the intensity of the Stark echo as a function of total delay.
  The dephasing mechanism in both cases is the same, the spread of
  frequencies in the original feature.  Therefore the FID provides an
  envelope for the amplitude of the Stark echoes.}
\end{figure}

FIG.~\ref{Stark_echo} shows the echoes created using a varying delay
between the input pulse and the electric field reversal.  The
intensity of the echo as a function of the delay is plotted in
FIG.~\ref{coherence}.  From this it can be seen that the envelope of
the echo amplitude has the same profile as the FID of the unbroadened
spectral feature.  The time-bandwidth product, or the number of
distinct pulses that can be stored is four.

\begin{figure}[!ht]
  \includegraphics[width=0.35\textwidth]{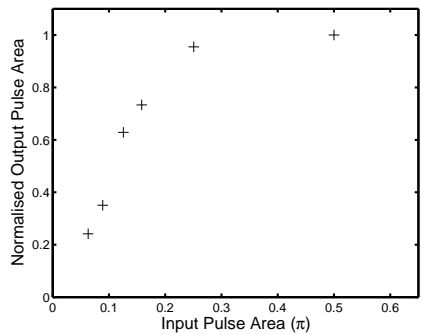}
  \caption{\label{attenuation}The energy of the Stark echo as a
    function of the area of the input pulse.  The response is linear
    until it saturates as the input pulse becomes comparable to a
    $\pi$/2 pulse.}
\end{figure}

The intensity of the echo for a 1.8~$\mu$s long input pulse as
a function of the input pulse intensity is shown in FIG.~\ref{attenuation}.  
At the highest input power of 7~mW the pulse area was $\pi$/2.  The 
size of the pulse corresponding to a $\pi$/2 pulse was determined by nutation 
measurements.  The
output amplitude is seen to be linear for low input intensities
but saturates for pulse areas approaching $\pi$/2.

In these initial experiments the efficiency of the echo was limited by
the low level of absorption of the broadened spectral feature. The
40\% absorption of the unbroadened line was reduced to 1\% on the
application of the electric field gradient.  For quantum memory
applications it will be necessary for the broadened feature to be
optically thick. 

One way to increase the optical thickness is to increase the
interaction length, using longer samples or multi-pass cells.
Alternately the absorption could be enhanced by placing the sample an
optical cavity.

Another way to increase the optical depth is by increasing the spectral
density of the ions.  With
doped samples it may not be possible to increase the spectral density
of the ions by simply increasing the dopant level \cite{sell04}. This is due to the
random strain in the crystal introduced by the dopant ions, producing
inhomogeneous broadening that increases linearly with the dopant
concentration \cite{cone}. An alternative strategy for achieving a
high spectral density is to use stoichiometric materials such as
EuCl$_{3}$.6H$_{2}$O.  The spectral density in EuCl$_{3}$.6H$_{2}$O is
high due to the high concentration of Eu$^{3+}$ ions and the low level
of strain induced broadening.



In conclusion we have shown that it is possible to rephase optical coherence
through reversing an external electric field gradient.  We 
demonstrate a time-bandwidth product of  four and linear operation 
at low input intensities. 
This rephasing using controlled inhomogeneous broadening is 
the first demonstration of 
this key effect which has potential applications as a quantum memory.  

An elegant aspect of this memory scheme, when compared to other
coherence optical memories, is that the only optical excitation is the
signal to be stored. The limitation of the current work was the low
optical thickness resulting in very low efficiency. Methods for
overcoming this limitation were discussed.

The authors would like to thank D. Freeman for his careful reading of
the manuscript. This work was supported by the Australian Defence
Science and Technology Organisation and the Australian Research Council.

\begin{thebibliography}{26}
\expandafter\ifx\csname natexlab\endcsname\relax\def\natexlab#1{#1}\fi
\expandafter\ifx\csname bibnamefont\endcsname\relax
  \def\bibnamefont#1{#1}\fi
\expandafter\ifx\csname bibfnamefont\endcsname\relax
  \def\bibfnamefont#1{#1}\fi
\expandafter\ifx\csname citenamefont\endcsname\relax
  \def\citenamefont#1{#1}\fi
\expandafter\ifx\csname url\endcsname\relax
  \def\url#1{\texttt{#1}}\fi
\expandafter\ifx\csname urlprefix\endcsname\relax\def\urlprefix{URL }\fi
\providecommand{\bibinfo}[2]{#2}
\providecommand{\eprint}[2][]{\url{#2}}

\bibitem[{\citenamefont{Cirac et~al.}(1997)\citenamefont{Cirac, Zoller, Kimble,
  and Mabuchi}}]{cira97}
\bibinfo{author}{\bibfnamefont{J.~I.} \bibnamefont{Cirac}},
  \bibinfo{author}{\bibfnamefont{P.}~\bibnamefont{Zoller}},
  \bibinfo{author}{\bibfnamefont{H.~J.} \bibnamefont{Kimble}},
  \bibnamefont{and} \bibinfo{author}{\bibfnamefont{H.}~\bibnamefont{Mabuchi}},
  \bibinfo{journal}{Phys. Rev. Lett.} \textbf{\bibinfo{volume}{78}},
  \bibinfo{pages}{3221} (\bibinfo{year}{1997}).

\bibitem[{\citenamefont{Fleischhauer and Lukin}(2000)}]{flei00}
\bibinfo{author}{\bibfnamefont{M.}~\bibnamefont{Fleischhauer}}
  \bibnamefont{and} \bibinfo{author}{\bibfnamefont{M.~D.} \bibnamefont{Lukin}},
  \bibinfo{journal}{Phys. Rev. Lett.} \textbf{\bibinfo{volume}{84}},
  \bibinfo{pages}{5094} (\bibinfo{year}{2000}).

\bibitem[{\citenamefont{Fleischhauer and Lukin}(2002)}]{flei02}
\bibinfo{author}{\bibfnamefont{M.}~\bibnamefont{Fleischhauer}}
  \bibnamefont{and} \bibinfo{author}{\bibfnamefont{M.~D.} \bibnamefont{Lukin}},
  \bibinfo{journal}{Phys. Rev. A} \textbf{\bibinfo{volume}{65}},
  \bibinfo{pages}{022314} (\bibinfo{year}{2002}).

\bibitem[{\citenamefont{Hau et~al.}(1999)\citenamefont{Hau, Harris, Dutton, and
  Behroozi}}]{hau99}
\bibinfo{author}{\bibfnamefont{L.~V.} \bibnamefont{Hau}},
  \bibinfo{author}{\bibfnamefont{S.~E.} \bibnamefont{Harris}},
  \bibinfo{author}{\bibfnamefont{Z.}~\bibnamefont{Dutton}}, \bibnamefont{and}
  \bibinfo{author}{\bibfnamefont{C.~H.} \bibnamefont{Behroozi}},
  \bibinfo{journal}{Nature} \textbf{\bibinfo{volume}{397}},
  \bibinfo{pages}{594} (\bibinfo{year}{1999}).

\bibitem[{\citenamefont{Liu et~al.}(2001)\citenamefont{Liu, Dutton, Behroozi,
  and Hau}}]{liu01}
\bibinfo{author}{\bibfnamefont{C.}~\bibnamefont{Liu}},
  \bibinfo{author}{\bibfnamefont{Z.}~\bibnamefont{Dutton}},
  \bibinfo{author}{\bibfnamefont{C.~H.} \bibnamefont{Behroozi}},
  \bibnamefont{and} \bibinfo{author}{\bibfnamefont{L.~V.} \bibnamefont{Hau}},
  \bibinfo{journal}{Nature} \textbf{\bibinfo{volume}{409}},
  \bibinfo{pages}{490} (\bibinfo{year}{2001}).

\bibitem[{\citenamefont{Kash et~al.}(1999)\citenamefont{Kash, Sautenkov,
  Zibrov, Hollberg, Welch, Rostovtsev, Fry, and Scully}}]{kash99}
\bibinfo{author}{\bibfnamefont{M.~M.} \bibnamefont{Kash}},
  \bibinfo{author}{\bibfnamefont{V.~A.} \bibnamefont{Sautenkov}},
  \bibinfo{author}{\bibfnamefont{A.~S.} \bibnamefont{Zibrov}},
  \bibinfo{author}{\bibfnamefont{L.}~\bibnamefont{Hollberg}},
  \bibinfo{author}{\bibfnamefont{G.~R.} \bibnamefont{Welch}},
  \bibinfo{author}{\bibfnamefont{M.~D. L.~Y.} \bibnamefont{Rostovtsev}},
  \bibinfo{author}{\bibfnamefont{E.~S.} \bibnamefont{Fry}}, \bibnamefont{and}
  \bibinfo{author}{\bibfnamefont{M.~O.} \bibnamefont{Scully}},
  \bibinfo{journal}{Phys. Rev. Lett.} \textbf{\bibinfo{volume}{82}},
  \bibinfo{pages}{5229} (\bibinfo{year}{1999}).

\bibitem[{\citenamefont{Budker et~al.}(1999)\citenamefont{Budker, Kimball,
  Rochester, and Yashchuk}}]{budk99}
\bibinfo{author}{\bibfnamefont{D.}~\bibnamefont{Budker}},
  \bibinfo{author}{\bibfnamefont{D.~F.} \bibnamefont{Kimball}},
  \bibinfo{author}{\bibfnamefont{S.~M.} \bibnamefont{Rochester}},
  \bibnamefont{and} \bibinfo{author}{\bibfnamefont{V.~V.}
  \bibnamefont{Yashchuk}}, \bibinfo{journal}{Phys. Rev. Lett.}
  \textbf{\bibinfo{volume}{83}}, \bibinfo{pages}{1767} (\bibinfo{year}{1999}).

\bibitem[{\citenamefont{Philips et~al.}(2001)\citenamefont{Philips,
  Fleischhauer, Mair, Walsworth, and Lukin}}]{phil01}
\bibinfo{author}{\bibfnamefont{D.~F.} \bibnamefont{Philips}},
  \bibinfo{author}{\bibfnamefont{A.}~\bibnamefont{Fleischhauer}},
  \bibinfo{author}{\bibfnamefont{A.}~\bibnamefont{Mair}},
  \bibinfo{author}{\bibfnamefont{R.~L.} \bibnamefont{Walsworth}},
  \bibnamefont{and} \bibinfo{author}{\bibfnamefont{M.~D.} \bibnamefont{Lukin}},
  \bibinfo{journal}{Phys. Rev. Lett.} \textbf{\bibinfo{volume}{86}},
  \bibinfo{pages}{783} (\bibinfo{year}{2001}).

\bibitem[{\citenamefont{Turukin et~al.}(2002)\citenamefont{Turukin,
  Sudarshanam, Shahriar, Musser, Ham, and Hemmer}}]{turu02}
\bibinfo{author}{\bibfnamefont{A.~V.} \bibnamefont{Turukin}},
  \bibinfo{author}{\bibfnamefont{V.~S.} \bibnamefont{Sudarshanam}},
  \bibinfo{author}{\bibfnamefont{M.~S.} \bibnamefont{Shahriar}},
  \bibinfo{author}{\bibfnamefont{J.~A.} \bibnamefont{Musser}},
  \bibinfo{author}{\bibfnamefont{B.~S.} \bibnamefont{Ham}}, \bibnamefont{and}
  \bibinfo{author}{\bibfnamefont{P.~R.} \bibnamefont{Hemmer}},
  \bibinfo{journal}{Phys. Rev. Lett.} \textbf{\bibinfo{volume}{88}},
  \bibinfo{pages}{023602} (\bibinfo{year}{2002}).

\bibitem[{\citenamefont{Longdell et~al.}(2005)\citenamefont{Longdell, Fraval,
  Sellars, and Manson}}]{long05}
\bibinfo{author}{\bibfnamefont{J.~J.} \bibnamefont{Longdell}},
  \bibinfo{author}{\bibfnamefont{E.}~\bibnamefont{Fraval}},
  \bibinfo{author}{\bibfnamefont{M.~J.} \bibnamefont{Sellars}},
  \bibnamefont{and} \bibinfo{author}{\bibfnamefont{N.~B.}
  \bibnamefont{Manson}}, \bibinfo{journal}{Phys. Rev. Lett.}
  \textbf{\bibinfo{volume}{95}}, \bibinfo{pages}{063601}
  (\bibinfo{year}{2005}).

\bibitem[{\citenamefont{Matsko et~al.}(2005)\citenamefont{Matsko, Strekalov,
  and Maleki}}]{mats05}
\bibinfo{author}{\bibfnamefont{A.~B.} \bibnamefont{Matsko}},
  \bibinfo{author}{\bibfnamefont{D.~V.} \bibnamefont{Strekalov}},
  \bibnamefont{and} \bibinfo{author}{\bibfnamefont{L.}~\bibnamefont{Maleki}},
  \bibinfo{journal}{Opt. Express} \textbf{\bibinfo{volume}{13}},
  \bibinfo{pages}{2210} (\bibinfo{year}{2005}).

\bibitem[{\citenamefont{Mossberg}(1982)}]{moss82}
\bibinfo{author}{\bibfnamefont{T.~W.} \bibnamefont{Mossberg}},
  \bibinfo{journal}{Opt. Lett.} \textbf{\bibinfo{volume}{7}},
  \bibinfo{pages}{77} (\bibinfo{year}{1982}).

\bibitem[{\citenamefont{Lin et~al.}(1995)\citenamefont{Lin, Wang, and
  Mossberg}}]{lin95}
\bibinfo{author}{\bibfnamefont{H.}~\bibnamefont{Lin}},
  \bibinfo{author}{\bibfnamefont{T.}~\bibnamefont{Wang}}, \bibnamefont{and}
  \bibinfo{author}{\bibfnamefont{T.~W.} \bibnamefont{Mossberg}},
  \bibinfo{journal}{Opt. Lett.} \textbf{\bibinfo{volume}{10}},
  \bibinfo{pages}{1658} (\bibinfo{year}{1995}).

\bibitem[{\citenamefont{Merkel et~al.}(2004)\citenamefont{Merkel, Mohan, Cole,
  Chang, Olson, and Babbit}}]{merk04}
\bibinfo{author}{\bibfnamefont{K.~D.} \bibnamefont{Merkel}},
  \bibinfo{author}{\bibfnamefont{R.~K.} \bibnamefont{Mohan}},
  \bibinfo{author}{\bibfnamefont{Z.}~\bibnamefont{Cole}},
  \bibinfo{author}{\bibfnamefont{T.}~\bibnamefont{Chang}},
  \bibinfo{author}{\bibfnamefont{A.}~\bibnamefont{Olson}}, \bibnamefont{and}
  \bibinfo{author}{\bibfnamefont{W.~R.} \bibnamefont{Babbit}},
  \bibinfo{journal}{J. Lumin} \textbf{\bibinfo{volume}{107}},
  \bibinfo{pages}{62} (\bibinfo{year}{2004}).

\bibitem[{\citenamefont{Moiseev and Kroll}(2001)}]{mois01}
\bibinfo{author}{\bibfnamefont{S.~A.} \bibnamefont{Moiseev}} \bibnamefont{and}
  \bibinfo{author}{\bibfnamefont{S.}~\bibnamefont{Kroll}},
  \bibinfo{journal}{Phys. Rev. Lett.} \textbf{\bibinfo{volume}{87}},
  \bibinfo{pages}{173601} (\bibinfo{year}{2001}).

\bibitem[{\citenamefont{Crisp}(1970)}]{cris70}
\bibinfo{author}{\bibfnamefont{M.~D.} \bibnamefont{Crisp}},
  \bibinfo{journal}{Phys. Rev. A} \textbf{\bibinfo{volume}{1}},
  \bibinfo{pages}{1604} (\bibinfo{year}{1970}).

\bibitem[{\citenamefont{Fraval et~al.}(2005)\citenamefont{Fraval, Sellars, and
  Longdell}}]{frav05}
\bibinfo{author}{\bibfnamefont{E.}~\bibnamefont{Fraval}},
  \bibinfo{author}{\bibfnamefont{M.~J.} \bibnamefont{Sellars}},
  \bibnamefont{and} \bibinfo{author}{\bibfnamefont{J.~J.}
  \bibnamefont{Longdell}}, \bibinfo{journal}{Phys. Rev. Lett.}
  \textbf{\bibinfo{volume}{95}}, \bibinfo{pages}{030506}
  (\bibinfo{year}{2005}).

\bibitem[{\citenamefont{Moiseev et~al.}(2003)\citenamefont{Moiseev, Tarasov,
  and Ham}}]{mois03}
\bibinfo{author}{\bibfnamefont{S.~A.} \bibnamefont{Moiseev}},
  \bibinfo{author}{\bibfnamefont{V.~F.} \bibnamefont{Tarasov}},
  \bibnamefont{and} \bibinfo{author}{\bibfnamefont{B.~S.} \bibnamefont{Ham}},
  \bibinfo{journal}{J. Opt. B} \textbf{\bibinfo{volume}{5}},
  \bibinfo{pages}{S497} (\bibinfo{year}{2003}).

\bibitem[{\citenamefont{Nilsson and Kroll}(2005)}]{nils05}
\bibinfo{author}{\bibfnamefont{M.}~\bibnamefont{Nilsson}} \bibnamefont{and}
  \bibinfo{author}{\bibfnamefont{S.}~\bibnamefont{Kroll}},
  \bibinfo{journal}{Opt. Comm.} \textbf{\bibinfo{volume}{247}},
  \bibinfo{pages}{393} (\bibinfo{year}{2005}).

\bibitem[{\citenamefont{Kraus et~al.}(2005)\citenamefont{Kraus, Tittel, Gisin,
  Nilsson, Kroll, and Cirac}}]{krau05}
\bibinfo{author}{\bibfnamefont{B.}~\bibnamefont{Kraus}},
  \bibinfo{author}{\bibfnamefont{W.}~\bibnamefont{Tittel}},
  \bibinfo{author}{\bibfnamefont{N.}~\bibnamefont{Gisin}},
  \bibinfo{author}{\bibfnamefont{M.}~\bibnamefont{Nilsson}},
  \bibinfo{author}{\bibfnamefont{S.}~\bibnamefont{Kroll}}, \bibnamefont{and}
  \bibinfo{author}{\bibfnamefont{J.}~\bibnamefont{Cirac}},
 (\bibinfo{year}{2005}),
  \eprint{quant-ph/0502184}.

\bibitem[{\citenamefont{Pryde et~al.}(2000)\citenamefont{Pryde, Sellars, and
  Manson}}]{pryd00}
\bibinfo{author}{\bibfnamefont{G.~J.} \bibnamefont{Pryde}},
  \bibinfo{author}{\bibfnamefont{M.~J.} \bibnamefont{Sellars}},
  \bibnamefont{and} \bibinfo{author}{\bibfnamefont{N.~B.}
  \bibnamefont{Manson}}, \bibinfo{journal}{Phys. Rev. Lett.}
  \textbf{\bibinfo{volume}{84}}, \bibinfo{pages}{1152} (\bibinfo{year}{2000}).

\bibitem[{\citenamefont{Longdell and Sellars}(2004)}]{tomog}
\bibinfo{author}{\bibfnamefont{J.~J.} \bibnamefont{Longdell}} \bibnamefont{and}
  \bibinfo{author}{\bibfnamefont{M.~J.} \bibnamefont{Sellars}},
  \bibinfo{journal}{Phys. Rev. A} \textbf{\bibinfo{volume}{69}},
  \bibinfo{pages}{032307} (\bibinfo{year}{2004}), \eprint{quant-ph/0208182}.

\bibitem[{\citenamefont{Longdell et~al.}(2004)\citenamefont{Longdell, Sellars,
  and Manson}}]{long04}
\bibinfo{author}{\bibfnamefont{J.~J.} \bibnamefont{Longdell}},
  \bibinfo{author}{\bibfnamefont{M.~J.} \bibnamefont{Sellars}},
  \bibnamefont{and} \bibinfo{author}{\bibfnamefont{N.~B.}
  \bibnamefont{Manson}}, \bibinfo{journal}{Phys. Rev. Lett.}
  \textbf{\bibinfo{volume}{93}}, \bibinfo{pages}{130503}
  (\bibinfo{year}{2004}).

\bibitem[{\citenamefont{Meixner et~al.}(1992)\citenamefont{Meixner, Jefferson,
  and Macfarlane}}]{meix92}
\bibinfo{author}{\bibfnamefont{A.~J.} \bibnamefont{Meixner}},
  \bibinfo{author}{\bibfnamefont{C.~M.} \bibnamefont{Jefferson}},
  \bibnamefont{and} \bibinfo{author}{\bibfnamefont{R.~M.}
  \bibnamefont{Macfarlane}}, \bibinfo{journal}{Phys. Rev. B}
  \textbf{\bibinfo{volume}{46}}, \bibinfo{pages}{5912} (\bibinfo{year}{1992}).

\bibitem[{\citenamefont{Sellars et~al.}(2004)\citenamefont{Sellars, Fraval, and
  Longdell}}]{sell04}
\bibinfo{author}{\bibfnamefont{M.~J.} \bibnamefont{Sellars}},
  \bibinfo{author}{\bibfnamefont{E.}~\bibnamefont{Fraval}}, \bibnamefont{and}
  \bibinfo{author}{\bibfnamefont{J.~J.} \bibnamefont{Longdell}},
  \bibinfo{journal}{J. Lumin} \textbf{\bibinfo{volume}{107}},
  \bibinfo{pages}{150} (\bibinfo{year}{2004}).

\bibitem[{\citenamefont{Cone}()}]{cone}
\bibinfo{author}{\bibfnamefont{R.~L.} \bibnamefont{Cone}},
  \emph{\bibinfo{title}{Personal communication}}.

\end{thebibliography}

\end{document}